\documentclass[10pt,conference,letterpaper]{IEEEtran}

\newif\ifonecolumn
\onecolumnfalse

\usepackage{pgf}
\usepackage{graphicx}
\graphicspath{{figures/}}
\usepackage{upgreek}
\usepackage{subfigure}
\usepackage{amsmath}
\usepackage{multicol}
\usepackage{multirow}
\usepackage{booktabs}
\usepackage{mathrsfs}
\usepackage{enumerate}
\usepackage{amssymb}
\usepackage{url}
\usepackage{bm}
\usepackage{stfloats}
\usepackage[hidelinks]{hyperref}
\usepackage{flushend}
\usepackage{epstopdf}
\usepackage{color}
\usepackage[mathscr]{euscript}
\usepackage{makecell}
\usepackage[square, comma, sort&compress, numbers]{natbib}
\usepackage{soul}

\newcommand{\xsfdel}[1]{\iffalse #1 \fi}

\begin{document}

\title{AB-Sync: Attention-Based Slot-Level Clock Synchronization Method for UWB-TDOA Localization Networks}

\author{Lyu Tianyi, Tian Kefei, Qin Kangqiao, Liu Qingwen, and Liu Mingqing}

\maketitle

\begin{abstract}
Ultra-wideband (UWB) time-difference-of-arrival (TDOA) localization networks provide high-update-rate indoor location services for IoT and cyber-physical applications, but their accuracy depends on nanosecond-level clock synchronization among anchors. Existing wireless clock synchronization (WCS) methods typically estimate clock states at the synchronization-stage or interval level, whereas TDMA-based UWB-TDOA systems localize tags from blinks transmitted in discrete short slots inside each synchronization stage. We identify this granularity mismatch as a source of residual TDOA error and present AB-Sync, an attention-based slot-level clock synchronization method. AB-Sync models the relationship between the slot-specific clock-speed ratio required by a target tag blink and neighboring clock-fluctuation observations, thereby enabling tag-slot-level timestamp mapping without adding extra UWB synchronization messages. On a real UWB-TDOA testbed, AB-Sync reduces the multi-anchor average TDOA ranging STD.V by 9.4\% and improves representative static localization accuracy by 18.6\% compared with Deferred+3S-KF, the leading low-overhead baseline in our evaluation. In a five-slot multi-tag experiment, AB-Sync consistently improves localization stability across all TDMA slots, reducing STD.V by 5.3\% on average and up to 16.2\% per slot with no extra UWB synchronization~overhead.
\end{abstract}

\section{Introduction}
\label{sec:Introduction}

Indoor localization is becoming a network service for industrial IoT, emergency response, asset tracking, and mobile cyber-physical systems. Recent wireless systems increasingly treat localization and spatial awareness as infrastructure-level capabilities, including indoor GNSS augmentation~\cite{wang2024gpms}, non-cooperative transmitter localization~\cite{lizarribar2024oransense}, high-dynamic tag tracking~\cite{harisha2025dragonfly}, smartphone UWB orientation sensing~\cite{zhou2024uwborient}, and UWB networks that scale toward many tags and dense mobile swarms~\cite{ma2025muloc,hou2026optimal}. These scenarios make both localization accuracy and channel occupancy critical. Among indoor radio technologies, ultra-wideband (UWB) is attractive because its sub-nanosecond timestamping resolution can support centimeter-level ranging. In UWB time-difference-of-arrival (TDOA) localization, a tag only broadcasts one blink, while multiple anchors timestamp its arrival time and recover the tag position from range-difference constraints. This one-transmission structure keeps the tag-side communication cost low, making TDOA suitable for dense multi-tag localization networks.

\begin{figure}[!t]
	\centering
    \includegraphics[scale=0.262]{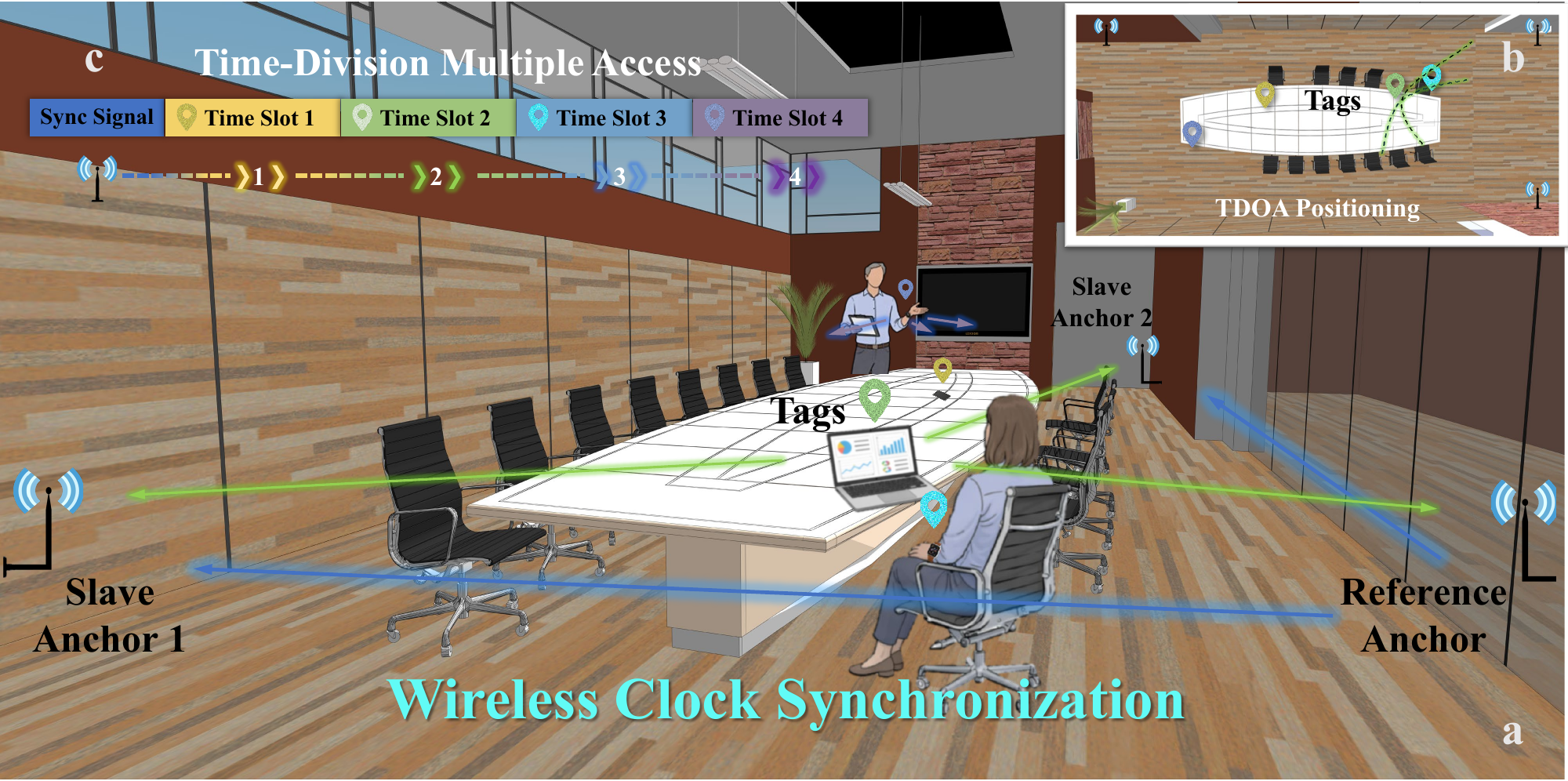}
	\caption{Low-overhead UWB-TDOA localization network. (a) The reference anchor broadcasts CCPs for wireless clock synchronization. (b) Anchors compute TDOA measurements for localization. (c) Tags transmit scheduled blinks in TDMA slots.}
    \label{fig:begining}
\end{figure}

The scalability of TDOA comes with a stringent synchronization requirement. Precise time synchronization is a network primitive for time-dependent distributed systems~\cite{finkenzeller2024ptpsec}; in UWB-TDOA localization, the dependency is even more direct. Since each anchor timestamps the same blink using its own clock, nanosecond-level clock mismatch can translate into tens of centimeters of ranging error. As shown in Fig.~\ref{fig:begining}(a) and Fig.~\ref{fig:begining}(b), practical UWB deployments therefore use wireless clock synchronization (WCS), where a reference anchor periodically broadcasts Clock Correction Packages (CCPs), and slave anchors estimate clock-deviation states for mapping their local timestamps onto the reference clock before TDOA localization. To preserve channel concurrency, low-overhead systems prefer unidirectional WCS: one CCP broadcast updates the synchronization state without requiring response packets from slave anchors.

Existing WCS methods mainly estimate clock states at the CCP-stage or interval level. A conventional real-time method uses previously observed CCP stages to predict the clock state for the next stage. Deferred synchronization improves this estimate by waiting until the following CCP is observed, so the clock state of the stage containing the tag blink can be estimated with lower uncertainty~\cite{Interpolation,Defer-3s-kf}. However, in a TDMA-based TDOA network, each localization measurement is triggered by a tag blink occurring at a specific short slot inside the CCP stage, as illustrated in Fig.~\ref{fig:begining}(c), whereas the WCS output is typically applied as a stage-level clock state. This creates a granularity mismatch: the clock state is estimated for a CCP stage, while the TDOA input is produced at a tag-slot instant within that stage. Assigning one stage-level clock state to all tag slots ignores residual intra-stage clock fluctuation, which becomes a bottleneck once prediction and filtering errors are reduced.

This paper addresses the granularity mismatch between stage-level clock-state estimation and slot-level TDOA measurement. Our key observation is that neighboring CCP-stage clock fluctuations contain information about the short-interval clock state at a specific tag slot, but the useful context is not fixed: historical and deferred observations may contribute differently depending on the slot position and local clock dynamics. This motivates a data-driven clock correction method that learns how to combine neighboring clock observations while keeping the UWB message overhead unchanged.

We propose AB-Sync, an attention-based slot-level clock-deviation correction method for low-overhead UWB-TDOA localization networks. AB-Sync does not introduce extra synchronization packets. Instead, it operates on host-collected CCP and blink timestamps, constructs a temporal neighborhood around the target CCP stage, and uses an attention-based model to estimate the clock-speed ratio required for the target tag slot. The corrected ratio is then used for TOA mapping and TDOA localization. In this way, AB-Sync refines synchronization from CCP-stage granularity to TDMA-slot granularity while preserving the one-message unidirectional synchronization structure.

The main contributions of this paper are summarized as follows.
\begin{itemize}
    \item We identify the granularity mismatch between CCP-stage-level clock-state estimation and TDMA slot-level tag transmissions in low-overhead UWB-TDOA localization networks. We show that stage-level clock correction leaves residual slot-level timing errors, which directly perturb the TDOA inputs used for localization.

    \item We propose AB-Sync, an attention-based slot-level clock-deviation correction method. AB-Sync uses neighboring clock-fluctuation observations before and after the target CCP stage to estimate the clock-speed ratio required for the target tag slot, while preserving the one-message unidirectional synchronization overhead of the baseline system.

    \item We implement AB-Sync on a real UWB-TDOA testbed and evaluate it from TDOA ranging to static 2D localization and multi-slot multi-tag operation. Compared with the leading low-overhead baseline, AB-Sync reduces the multi-anchor average TDOA ranging STD.V by 9.4\%, improves representative static localization accuracy by 18.6\%, and achieves consistent localization gains across five TDMA slots.
\end{itemize}

\begin{figure}[!t]
	\centering
    \includegraphics[width=\columnwidth]{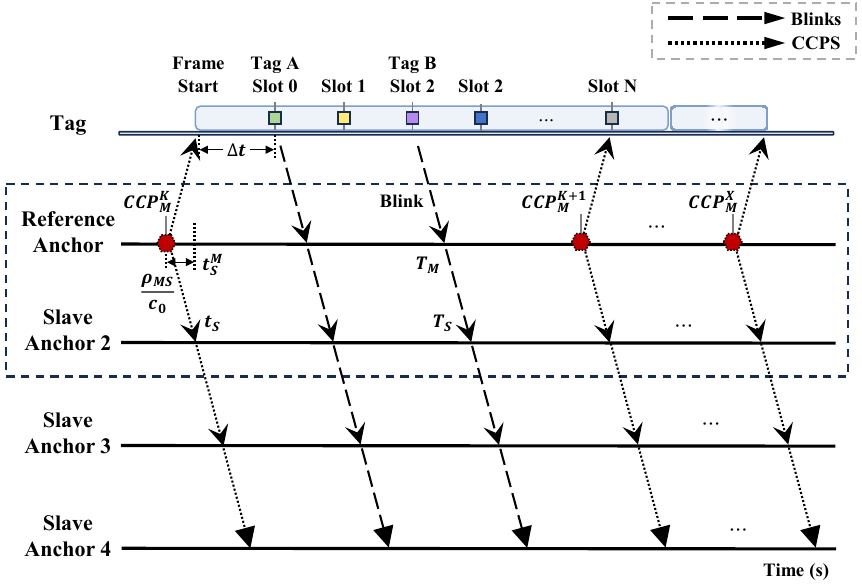}
	\caption{Timing structure of wireless clock synchronization and TDMA tag blinks in a UWB-TDOA localization network.}
    \label{fig:synchronization scheme}
\end{figure}

\section{System Model and Problem Statement}\label{sec:system_model}

\subsection{UWB-TDOA Timing Model}\label{subsection:uwb_tdoa_timing_model}

We consider an uplink UWB-TDOA localization network consisting of one reference anchor, multiple slave anchors, and TDMA-scheduled tags. As shown in Fig.~\ref{fig:synchronization scheme}, the reference anchor periodically broadcasts Clock Correction Packages (CCPs), while tags transmit blinks in scheduled TDMA slots. Slave anchors timestamp both CCPs and tag blinks using their local clocks, and the host maps slave-anchor timestamps to the reference-anchor clock before computing TDOA values.

Let the reference-clock transmission time of the $k$-th CCP be $\mathrm{CCP}_M^k$, and let its reception time recorded by a slave anchor be $t_S[k]$. We use the superscript $M$ to denote a timestamp measured on, or mapped to, the reference-anchor clock. Thus, $t_S^M[k]$ denotes the arrival time of the same CCP at the slave anchor expressed on the reference clock. Since the anchor geometry is known, this reference-clock arrival time is
\begin{equation}\label{formula:tSM}
t_S^M[k]
=
\mathrm{CCP}_M^k+\frac{\rho^{MS}}{c_0},
\end{equation}
where $\rho^{MS}$ is the distance between the reference and slave anchors and $c_0$ is the speed of light. The clock offset between the slave clock and the reference clock at the CCP arrival moment is therefore
\begin{equation}\label{formula:offset}
\theta_S^M[k]=t_S^M[k]-t_S[k].
\end{equation}

\begin{figure}[!t]
	\centering
    \includegraphics[width=0.82\columnwidth]{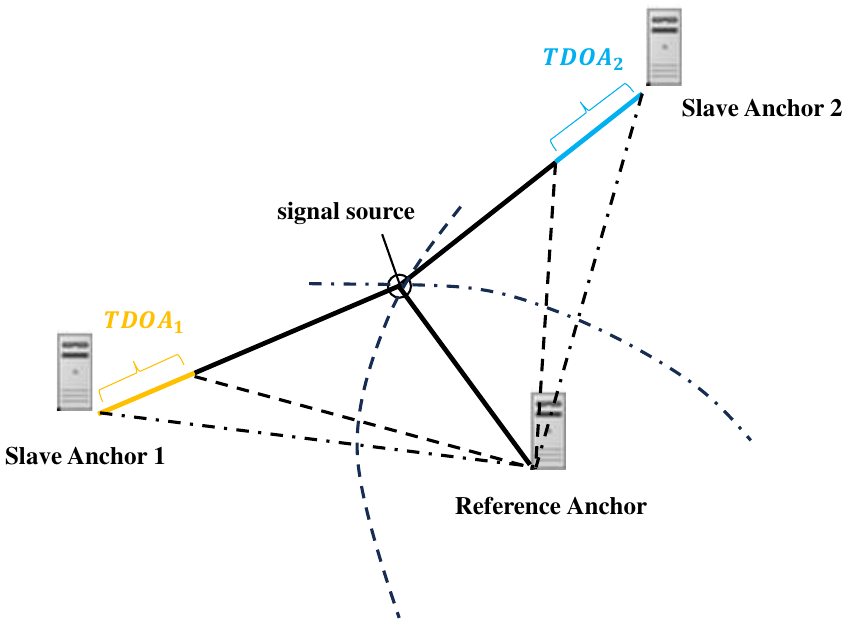}
	\caption{Three-anchor TDOA positioning geometry.}
    \label{fig:TDOA_algorithm}
\end{figure}

The clock-speed ratio over the completed CCP stage $[k-1,k]$ is estimated from the CCP sending interval measured by the reference anchor and the reception interval measured by the slave anchor:
\begin{equation}\label{formula:traditional_Ratio}
\mathrm{Ratio}_{[k-1]\sim[k]}
=
\frac{\mathrm{Syn}_M^{[k-1]\sim[k]}}
{\mathrm{Syn}_S^{[k-1]\sim[k]}} .
\end{equation}
where $\mathrm{Syn}_M^{[k-1]\sim[k]}$ and $\mathrm{Syn}_S^{[k-1]\sim[k]}$ denote the elapsed time between the $(k-1)$-th and $k$-th CCPs measured by the reference clock and the slave clock, respectively. This ratio is used to map a slave-clock interval to the reference clock. For a tag blink received after the $k$-th CCP, the mapped slave-anchor TOA can be written as
\begin{equation}\label{deltaTM}
T_S^M
=
\mathrm{CCP}_M^k+\frac{\rho^{MS}}{c_0}+\Delta T_M,
\qquad
\Delta T_M \approx \Delta T_S \cdot r ,
\end{equation}
where $T_S^M$ is the slave-anchor blink TOA after being mapped to the reference clock, $\Delta T_S$ is the interval from CCP reception to blink reception measured by the slave clock, $\Delta T_M$ is the corresponding reference-clock interval, and $r$ is the clock-speed ratio used for TOA mapping. The TDOA measurement is then computed as $\mathrm{TDOA}=T_S^M-T_M$, where $T_M$ is the blink TOA recorded by the reference anchor. As shown in Fig.~\ref{fig:TDOA_algorithm}, a two-dimensional TDOA solver uses at least three anchors. If $d_1$ is the tag distance to the reference anchor and $d_2,d_3$ are the distances to two slave anchors, the two range-difference constraints are
\begin{equation}
\begin{aligned}
d_2-d_1 &= c_0 \cdot (T_{S1}^M-T_M),\\
d_3-d_1 &= c_0 \cdot (T_{S2}^M-T_M).
\end{aligned}
\label{formula:tdoa_geometry}
\end{equation}

These hyperbolic constraints connect clock synchronization to localization: reducing the fluctuation of mapped TOAs directly reduces the uncertainty of the TDOA inputs used by the positioning solver.

\subsection{Stage-to-Slot Granularity Mismatch}\label{subsection:stage_slot_mismatch}

The key synchronization question is which ratio should be used as $r$ for the tag blink. Existing WCS methods mainly differ in whether this ratio is predicted from previous CCP stages or estimated after a deferred CCP becomes available:
\begin{equation}\label{formula:wcs_ratio_granularity}
r=
\begin{cases}
\mathrm{Ratio}_{[k-1]\sim[k]}, & \text{real-time stage-level mapping},\\
\mathrm{Ratio}_{[k]\sim[k+1]}, & \text{deferred stage-level mapping},\\
\mathrm{Ratio}_{[k,k+\Delta T_S]}, & \text{ideal slot-level mapping}.
\end{cases}
\end{equation}
Conventional real-time mapping supports immediate correction but uses a clock state observed before the blink stage. Deferred mapping, such as LI and Deferred+3S-KF~\cite{Interpolation,Defer-3s-kf}, waits until the following CCP is observed and therefore obtains a lower-uncertainty estimate of the stage containing the blink. More advanced methods, such as 2S-KF and 3S-KF, do not simply use the raw stage-wise ratio. Instead, they behave as history-dependent filters that recursively estimate a higher-order clock-state vector from a sequence of CCP observations:
\begin{equation}
\widehat{\mathbf{x}}_k^{\mathrm{WCS}}
=
\mathcal{F}_{\mathrm{WCS}}\!\left(\mathcal{C}_{1:k+\ell}\right),
\qquad
r_k^{\mathrm{WCS}}
=
h\!\left(\widehat{\mathbf{x}}_k^{\mathrm{WCS}}\right),
\label{formula:generic_wcs_state}
\end{equation}
where $\widehat{\mathbf{x}}_k^{\mathrm{WCS}}$ is the clock state estimated for the target CCP stage, $\mathcal{C}_{1:k+\ell}$ denotes the CCP observation sequence from the first stage to stage $k+\ell$, and $\mathcal{F}_{\mathrm{WCS}}(\cdot)$ denotes the corresponding filtering or synchronization algorithm. The parameter $\ell$ indicates whether future CCP observations are used by a deferred method, and $h(\cdot)$ extracts the clock-speed ratio used for TOA mapping. For example, 3S-KF tracks higher-order clock deviations such as offset, skew, and skew variation, while Deferred+3S-KF further updates this state after observing the following CCP. Nevertheless, the resulting $r_k^{\mathrm{WCS}}$ is still a stage-level quantity applied to all tag blinks inside the stage.

As shown in Fig.~\ref{fig:synchronization scheme}, each tag blink in a TDMA-based TDOA network occupies a short slot within the CCP stage. The actual synchronization target is therefore the short-interval ratio $\mathrm{Ratio}_{[k,k+\Delta T_S]}$, not merely the average ratio over the full CCP stage. This stage-to-slot granularity mismatch motivates AB-Sync to estimate a slot-level clock-deviation correction from neighboring CCP-stage observations while preserving the low-overhead unidirectional synchronization structure.

\subsection{Slot-Level Correction Objective}\label{subsection:slot_level_objective}

The above mismatch can be written as a timestamp-mapping error. Let $s=(k,\Delta T_S)$ denote a tag blink received in the target slot after the $k$-th CCP, and let $r_s^\star=\mathrm{Ratio}_{[k,k+\Delta T_S]}$ be the ideal short-interval clock-speed ratio for this slot. If a WCS method uses a stage-level ratio $r_k^{\mathrm{WCS}}$ for this blink, the residual TOA mapping error at the slave anchor is
\begin{equation}
e_s^T
=
\Delta T_S \cdot \left(r_k^{\mathrm{WCS}}-r_s^\star\right).
\label{formula:slot_mapping_error}
\end{equation}
where $e_s^T$ denotes the TOA-domain mapping error of slot $s$. The corresponding range-difference error is proportional to $c_0 \cdot e_s^T$. Therefore, even when the stage-level WCS ratio is accurate on average, a small ratio mismatch can still introduce a non-negligible TDOA error if it occurs at the exact tag-transmission slot.

AB-Sync formulates synchronization as a slot-level regression problem. Given only the CCP observations already collected for unidirectional WCS and the blink timestamp metadata already collected for localization, AB-Sync learns a mapping
\begin{equation}
\widetilde{r}_s
=
g_\theta\!\left(\mathcal{C}_{k-L_p:k+L_d},\Delta T_S\right),
\label{formula:slot_regression_objective}
\end{equation}
where $\widetilde{r}_s$ is the AB-Sync-estimated slot-level ratio, $g_\theta(\cdot)$ is the learned regression model with trainable parameters $\theta$, and $\mathcal{C}_{k-L_p:k+L_d}$ denotes the neighboring CCP-stage clock observations visible around the target stage. Here, $L_p$ and $L_d$ are the numbers of preceding and deferred CCP stages included in the observation window. The goal is not to replace the low-overhead WCS protocol, but to refine the ratio used by TOA mapping from a stage-level quantity $r_k^{\mathrm{WCS}}$ to a slot-level estimate $\widetilde{r}_s$:
\begin{equation}
\widetilde{r}_s
\approx
r_s^\star,
\qquad
\Delta T_S \cdot \widetilde{r}_s
\approx
\Delta T_S \cdot r_s^\star .
\label{formula:slot_level_goal}
\end{equation}
where the second relation states the equivalent requirement in the mapped reference-clock interval. This objective also clarifies the deployment constraint of AB-Sync: improving slot-level synchronization must not consume additional UWB channel time. All learning and inference are performed on the host side using timestamps that are already available in a standard UWB-TDOA localization pipeline.

\begin{figure*}[!t]
    \centering
    \includegraphics[width=\textwidth]{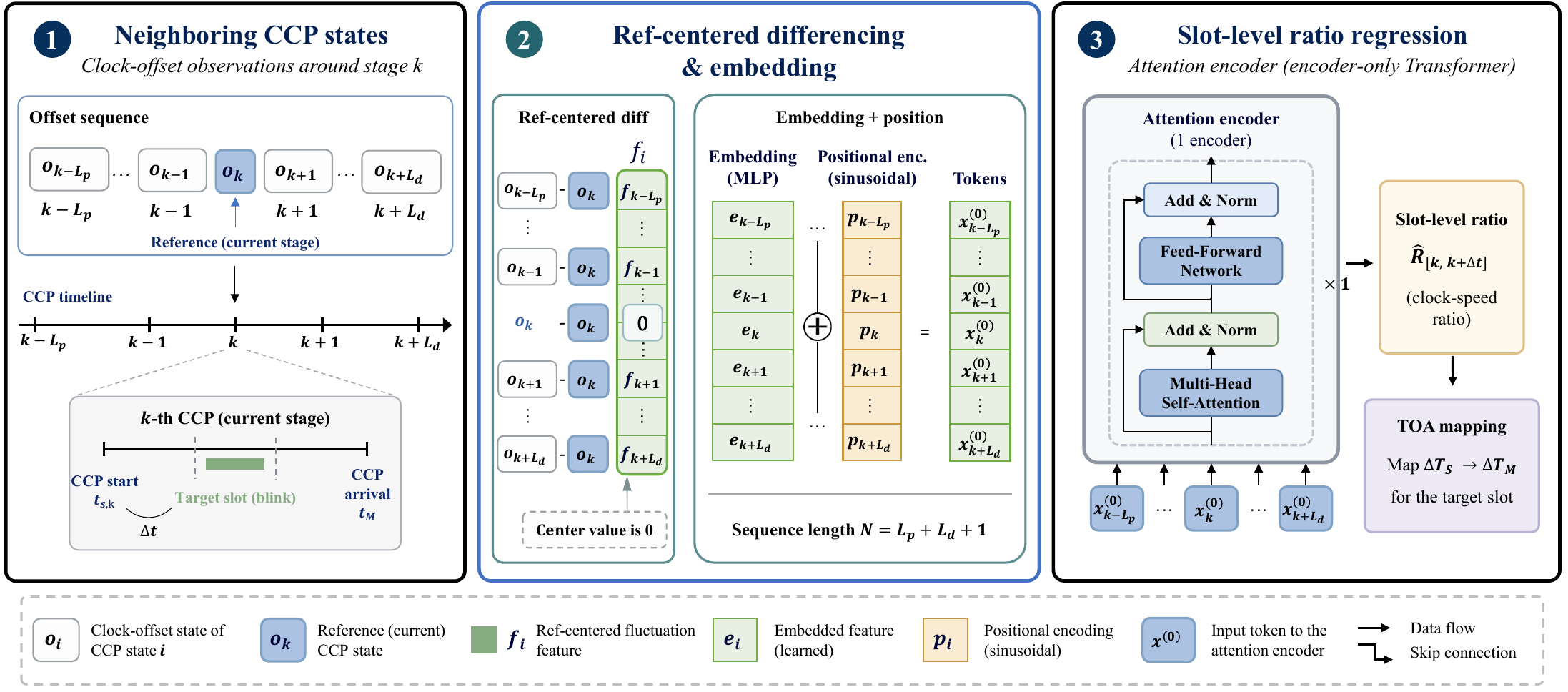}
    \caption{Pipeline of the proposed AB-Sync method, which extracts neighboring CCP-stage clock observations, converts them into ref-centered fluctuation features, and regresses the slot-level clock-speed ratio for TOA mapping through an attention-based network.}
    \label{fig:ab_sync_pipeline}
\end{figure*}

\section{AB-Sync Design}\label{sec:absync_design}

\subsection{Slot-Level WCS Formulation and Attention Model}\label{subsection:proposed_attention_wcs}

AB-Sync estimates the slot-level ratio $\widetilde{r}_s\triangleq\widetilde{\mathrm{Ratio}}_{[k,k+\Delta T_S]}$ for a tag slot $s=(k,\Delta T_S)$, which approximates the ideal short-interval ratio $r_s^\star=\mathrm{Ratio}_{[k,k+\Delta T_S]}$ defined in Section~\ref{subsection:slot_level_objective}. For a blink transmitted between the $k$-th and $(k+1)$-th CCPs, the slot-level mapping is
\begin{equation}
\begin{aligned}
T_S^M
=
\mathrm{CCP}_M^k
+\frac{\rho^{MS}}{c_0}
+\Delta T_S \cdot \widetilde{r}_s .
\end{aligned}
\label{formula:attn_slot_mapping}
\end{equation}

The central design question is how to infer the slot-level ratio from CCP-stage observations. Huang et al.~\cite{2S_KF_deferred} adopt a fixed interpolation rule that assigns the contribution of neighboring CCP stages according to their temporal distance from the target tag slot. Under this rule, a tag transmitted near the beginning of a CCP stage may be more related to the preceding stage, while a tag transmitted near the end may benefit more from deferred observations. However, this time-distance-driven rule is still rigid: local oscillator fluctuation, packet timestamp noise, and propagation-path-induced measurement fluctuation can make the useful temporal context asymmetric. AB-Sync therefore treats slot-level synchronization as dynamic temporal weighting: the model learns which neighboring CCP-stage observations should contribute to the target slot under the current clock-fluctuation pattern. Figure~\ref{fig:ab_sync_pipeline} summarizes this AB-Sync inference pipeline, from neighboring CCP-stage observation extraction and ref-centered fluctuation construction to attention-based temporal weighting and slot-level clock-speed ratio correction.

The model uses a temporal neighborhood around the tag-blink stage located between the $k$-th and $(k+1)$-th CCPs. With $L_p$ preceding stage-interval states and $L_d$ deferred stage-interval states, the visible window and its offset-state sequence are
\begin{equation}
\begin{aligned}
\mathcal{N}_k&=\{k-L_p,\ldots,k,\ldots,k+L_d\},\\
L&=|\mathcal{N}_k|=L_p+L_d+1,
\end{aligned}
\label{formula:attn_neighborhood}
\end{equation}
\begin{equation}
\mathbf{O}_k=[o_i]_{i\in\mathcal{N}_k}^{T},
\label{formula:offset_sequence}
\end{equation}
where $\mathcal{N}_k$ is the index set of visible CCP stage intervals around the tag-blink stage $[k,k+1]$, $L$ is the sequence length, $\mathbf{O}_k$ is the offset-state sequence, and $o_i$ is the clock-offset state of the completed interval between the $i$-th and $(i+1)$-th CCPs. To emphasize the local fluctuation pattern rather than the absolute offset magnitude, AB-Sync applies ref-centered differencing:
\begin{equation}
\mathbf{F}_k=[o_i-o_k]_{i\in\mathcal{N}_k}^{T}.
\label{formula:ref_diff_feature}
\end{equation}
where $\mathbf{F}_k$ is the input fluctuation sequence and $o_k$ is the offset state of the blink-containing interval $[k,k+1]$. This ref-centered representation removes the large deterministic offset component and keeps the local clock-deviation pattern that is most relevant to short-interval correction. In other words, AB-Sync learns from the shape of neighboring clock fluctuations rather than from an anchor-pair-specific offset magnitude.

The fluctuation sequence is then embedded and combined with standard sinusoidal positional encoding before entering an encoder-only Transformer:
\begin{equation}
\begin{aligned}
\mathbf{X}^{(0)}&=\mathrm{Embed}(\mathbf{F}_k)+\mathbf{P},\\
\bar{\mathbf{X}}&=\mathrm{TransformerEncoder}(\mathbf{X}^{(0)}).
\end{aligned}
\label{formula:encoder_input_matrix}
\end{equation}
where $\mathrm{Embed}(\cdot)$ maps the scalar fluctuation sequence to hidden vectors, $\mathbf{P}$ is the positional encoding, $\mathbf{X}^{(0)}$ is the encoder input sequence, and $\bar{\mathbf{X}}$ is the encoder output sequence. Inside the encoder, multi-head self-attention allows each CCP-stage feature to be recombined with its neighboring stages. The attention operation is the standard scaled dot-product form:
\begin{equation}
\mathrm{Attn}(\mathbf{Q},\mathbf{K},\mathbf{V})
=
\mathrm{softmax}\!\left(
\frac{\mathbf{Q}\mathbf{K}^{T}}{\sqrt{d_h}}
\right)\mathbf{V}.
\label{formula:attention_weight}
\end{equation}
where $\mathbf{Q}$, $\mathbf{K}$, and $\mathbf{V}$ are the query, key, and value matrices obtained from the embedded clock-fluctuation sequence, and $d_h$ is the dimension of each attention head. This compact notation keeps the design focused on the temporal clock-regression problem; the remaining layer-normalization, residual, and feed-forward operations follow the standard Transformer encoder.

Since the fluctuation sequence is ordered by increasing interval index, the $(L_p+1)$-th sequence element corresponds to the reference interval around the target slot. AB-Sync uses the corresponding encoder output $\bar{\mathbf{x}}_{L_p+1}$ as the readout vector for slot-level correction. The prediction head outputs the centered offset fluctuation for slot $s$:
\begin{equation}
\widetilde{f}_s
=
\mathbf{w}_p^T\bar{\mathbf{x}}_{L_p+1}+b_p .
\label{formula:attention_residual}
\end{equation}
where $\mathbf{w}_p$ and $b_p$ are the weight vector and bias of the prediction head. The predicted value $\widetilde{f}_s$ represents the centered offset-fluctuation estimate for the target slot. After adding back the reference offset $o_k$, the corrected slot-level clock-speed ratio is recovered as
\begin{equation}
\widetilde{r}_s
=
\frac{S_M}{S_M+o_k+\widetilde{f}_s} ,
\label{formula:attention_delta}
\end{equation}
where $S_M=\mathrm{Syn}_M^{[k]\sim[k+1]}$ is the reference-clock interval of the blink-containing CCP stage. The recovered ratio is applied to the blink-specific interval $\Delta T_S$ in (\ref{formula:attn_slot_mapping}).

\vspace{4pt}
\textbf{Supervision Label Derivation:} During training, a tag with known geometry provides the supervised slot-level ratio because its ground-truth TDOA can be computed from the anchor and tag positions. For compactness, we use the same slot index $s=(k,\Delta T_S)$. The supervised ratio is
\begin{equation}
\widehat{r}_{s}
=
\frac{
T_M+\widehat{\mathrm{TDOA}}_{s}
-\mathrm{CCP}_M^k-\widehat{\rho}^{MS}/c_0
}
{\Delta T_S}.
\label{formula:attn_ratio_label}
\end{equation}
where $\widehat{r}_s$ is the supervised slot-level ratio, $\widehat{\mathrm{TDOA}}_{s}$ is the geometry-derived ground-truth TDOA of slot $s$, and $\widehat{\rho}^{MS}$ is the calibrated distance between the reference and slave anchors. For consistency with the network output, this ratio is converted into the centered offset-fluctuation target:
\begin{equation}
\widehat{f}_{s}
=
\frac{S_M}{\widehat{r}_{s}}-S_M-o_k .
\label{formula:attention_target}
\end{equation}
The model is trained by minimizing the mini-batch MSE between the predicted and supervised centered fluctuation:
\begin{equation}
\mathcal{L}(\theta)
=
\frac{1}{N_b}\sum_{n=1}^{N_b}
\left(
\widetilde{f}_{n}-\widehat{f}_{n}
\right)^2 .
\label{formula:attention_loss}
\end{equation}
where $N_b$ is the mini-batch size, and $\widetilde{f}_{n}$ and $\widehat{f}_{n}$ are the predicted and supervised centered fluctuation values of the $n$-th training sample. Thus, AB-Sync ultimately learns to infer a short-interval clock state from neighboring CCP-stage clock fluctuations and outputs the corrected slot-level ratio for TDOA calculation.

\section{Evaluation}\label{sec:evaluation}

\subsection{Experimental Setup}\label{subsection:experimental_setup}

We evaluate AB-Sync on a real UWB-TDOA testbed built with DWM1000-based anchors and tags. The reference anchor periodically broadcasts CCP messages, while slave anchors timestamp both CCPs and tag blinks using their local clocks. The anchors upload raw CCP and blink timestamp metadata to a host computer, where AB-Sync constructs the neighboring CCP-stage observation window, regresses the slot-level clock-speed ratio, maps the slave-anchor TOA to the reference clock, and computes TDOA values for localization. This design preserves the one-message unidirectional synchronization structure: AB-Sync does not introduce additional UWB synchronization packets and only changes the host-side clock-state correction used for TOA mapping.

\begin{figure}[!t]
    \centering
    \includegraphics[scale=0.38]{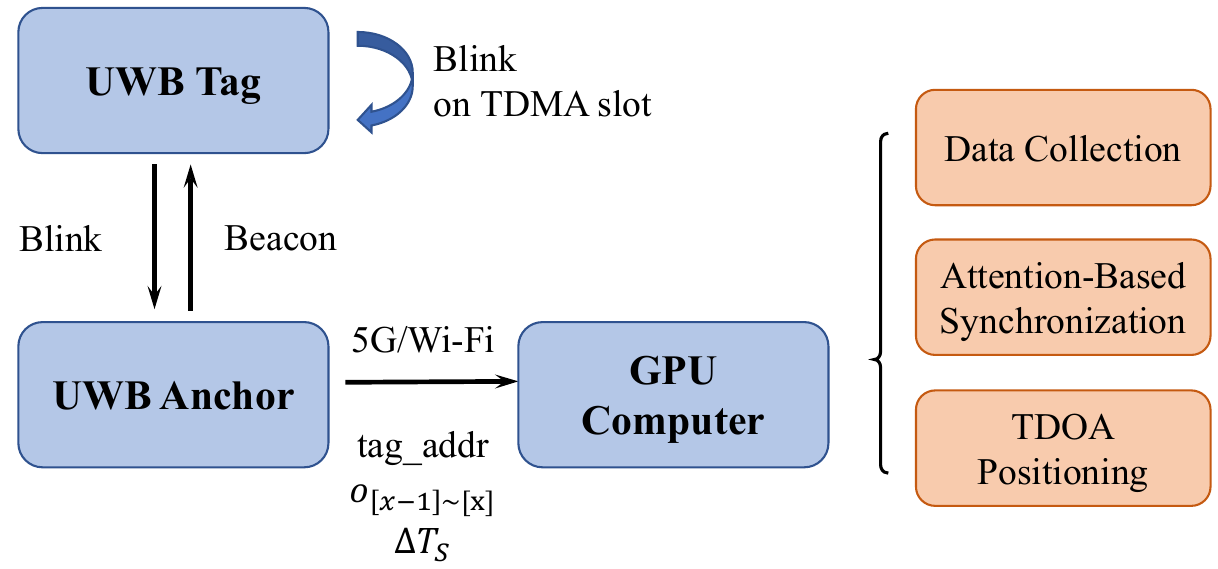}
    \caption{System architecture of the UWB-TDOA testbed with host-side AB-Sync clock-state correction.}
    \label{fig:workflow}
\end{figure}

Table~\ref{tab:ab_sync_training_configuration} summarizes the common testbed, data-splitting, and evaluation protocol followed by all experiments in this section unless explicitly stated otherwise. The evaluation first uses a single fixed 50~ms tag slot to study context-window selection, the corresponding TDOA ranging stability, and representative static 2D localization in Subsections~IV-B--IV-D, and then extends the same protocol to five TDMA tag-transmission slots in Subsection~IV-E.

\begin{table}[!t]
\centering
\caption{Experimental and training setup shared by the AB-Sync experiments.}
\label{tab:ab_sync_training_configuration}
\renewcommand{\arraystretch}{1.08}
\setlength{\tabcolsep}{3pt}
\resizebox{\columnwidth}{!}{%
\begin{tabular}{ll}
\toprule
\textbf{Item} & \textbf{Setting} \\
\midrule
UWB module & DWM1000-based anchors and tags \\
Channel / data rate & Channel~2, 3993.6~MHz / 110~kbps \\
Synchronization mode & One CCP broadcast per stage \\
CCP / blink period & 100~ms / 400~ms \\
Single-slot study & Blink scheduled at the 50~ms slot \\
Multi-slot study & Five TDMA tag-transmission slots \\
Dataset split & 55\% train / 5\% validation / 40\% test \\
AB-Sync window & Selected by $L_p$/$L_d$ sweep \\
Optimizer / loss & Adam / MSE \\
Baselines & Conv., Deferred, Conv.+3S-KF, Deferred+3S-KF \\
Metrics & TDOA and 2D localization RMSE/ME/STD.V \\
\bottomrule
\end{tabular}%
}
\end{table}

\subsection{Context-Window Selection}\label{subsection:context_window_selection}

This subsection investigates the optimal clock-deviation-state context window used by AB-Sync to regress the clock state of a short tag-arrival interval. This provides insight into how many neighboring CCP stages are informative for estimating the clock state associated with a short interval inside one CCP stage.

Following the timing and geometry models defined in Section~\ref{sec:system_model}, this subsection evaluates how the visible temporal neighborhood affects AB-Sync before fixing the window used by later experiments. For a static target, the ground-truth TDOA remains constant over consecutive blinks; therefore, a more stable TDOA sequence indicates a more accurate and stable clock synchronization result.

In this experiment, the reference anchor and three slave anchors were deployed at $\mathbf{r}_1=(0,0)$, $\mathbf{r}_2=(9.3,9.3)$, $\mathbf{r}_3=(3.9,8.7)$, and $\mathbf{r}_4=(2.4,0)$, respectively, with coordinates measured in meters. The tag was placed statically at $\mathbf{r}_T=(3.6,6.7)$, and all CCP/blink timestamps were processed by the host-side pipeline described in Subsection~\ref{subsection:experimental_setup}.

For the context-window study, one AB-Sync model was trained for each combination of $L_p$ and $L_d$ and for each anchor pair. The tag transmitted one blink every 400~ms. To obtain a typical and fixed short interval inside each CCP stage, the blink was scheduled at the 50~ms slot, i.e., the midpoint of each 100~ms CCP stage. After removing incomplete samples, the available sample counts for Anchors~1\&2, Anchors~1\&3, and Anchors~1\&4 were 8775, 8788, and 8789, respectively. The samples were split chronologically into training, validation, and test sets with a ratio of 55\%, 5\%, and 40\%, respectively. The shared experimental and training settings are summarized in Table~\ref{tab:ab_sync_training_configuration}.

Table~\ref{tab:ab_sync_context_window} summarizes the effect of the visible temporal neighborhood size on TDOA ranging stability. The full sweep also includes the boundary case $L_d=0$; to keep the table compact, Table~\ref{tab:ab_sync_context_window} reports the more representative deferred-window settings with $L_d=1,\ldots,4$. Each entry reports the STD.V results of three anchor pairs in the form of Anchors~1\&2 / Anchors~1\&3 / Anchors~1\&4, which follows the order used in the table. For each anchor pair, the lowest STD.V is highlighted in bold, and the second-lowest STD.V is underlined. Here, $L_p$ and $L_d$ follow the preceding/deferred window convention defined in Section~III. The results show that increasing the visible neighborhood generally improves the TDOA stability, while an excessively large deferred window does not always bring further improvement. This suggests that CCP-stage observations within about 300~ms before and after the target interval usually capture the most relevant short-term clock variations for slot-level correction. In this experiment, $L_p=3$ and $L_d=3$ provide a balanced setting across the three anchor pairs and are therefore adopted in the subsequent TDOA and localization evaluations.

\begin{table}[!t]
\centering
\caption{Effect of visible temporal neighborhood size on TDOA ranging accuracy (STD.V, cm). Each entry is reported as Anchors~1\&2 / Anchors~1\&3 / Anchors~1\&4.}
\label{tab:ab_sync_context_window}
\renewcommand{\arraystretch}{1.08}
\setlength{\tabcolsep}{2pt}
\resizebox{0.95\columnwidth}{!}{%
\begin{tabular}{@{}ccccc@{}}
\toprule
$L_p$\textbackslash$L_d$ & 1 & 2 & 3 & 4 \\
\midrule
0 & 9.47/6.10/5.66 & 8.56/5.65/\textbf{5.54} & 8.44/5.62/5.57 & \textbf{8.26}/5.62/5.60 \\
1 & 9.45/6.09/5.64 & 8.44/5.67/5.58 & 8.37/\textbf{5.58}/5.57 & 8.60/\textbf{5.58}/5.59 \\
2 & 9.36/6.05/5.63 & 8.49/5.67/\textbf{5.54} & 8.46/5.61/5.58 & 8.52/5.87/5.57 \\
3 & 9.24/6.02/5.62 & 8.45/\underline{5.59}/5.56 & \underline{8.31}/\underline{5.59}/\underline{5.55} & 8.42/5.62/5.58 \\
4 & 9.20/5.98/5.62 & 8.64/5.70/5.56 & 8.65/5.67/5.57 & 8.68/5.67/5.58 \\
\bottomrule
\end{tabular}%
}
\end{table}

\subsection{TDOA Ranging Accuracy}\label{subsection experiment tdoa range}

\begin{figure*}[t]
    \centering
    \begin{minipage}[t]{0.32\textwidth}
        \centering
        \includegraphics[page=3,width=\textwidth]{tdoa_ranging_time_distribution_comparison.pdf}
        \parbox{\textwidth}{\centering (a)}
    \end{minipage}
    \hfill
    \begin{minipage}[t]{0.32\textwidth}
        \centering
        \includegraphics[page=2,width=\textwidth]{tdoa_ranging_time_distribution_comparison.pdf}
        \parbox{\textwidth}{\centering (b)}
    \end{minipage}
    \hfill
    \begin{minipage}[t]{0.32\textwidth}
        \centering
        \includegraphics[page=1,width=\textwidth]{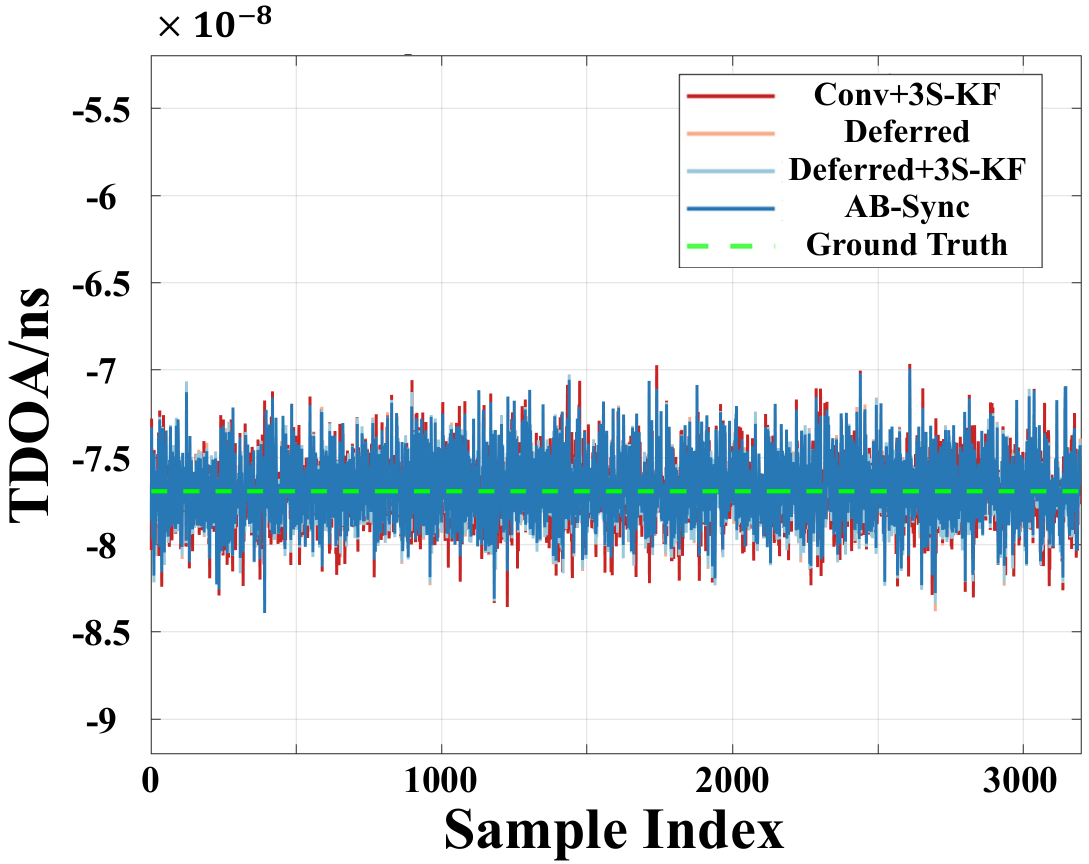}
        \parbox{\textwidth}{\centering (c)}
    \end{minipage}

    \vspace{3pt}

    \begin{minipage}[t]{0.32\textwidth}
        \centering
        \includegraphics[page=6,width=\textwidth]{tdoa_ranging_time_distribution_comparison.pdf}
        \parbox{\textwidth}{\centering (d)}
    \end{minipage}
    \hfill
    \begin{minipage}[t]{0.32\textwidth}
        \centering
        \includegraphics[page=5,width=\textwidth]{tdoa_ranging_time_distribution_comparison.pdf}
        \parbox{\textwidth}{\centering (e)}
    \end{minipage}
    \hfill
    \begin{minipage}[t]{0.32\textwidth}
        \centering
        \includegraphics[page=4,width=\textwidth]{tdoa_ranging_time_distribution_comparison.pdf}
        \parbox{\textwidth}{\centering (f)}
    \end{minipage}

\caption{TDOA time-domain sequences and probability-density distributions for different anchor pairs using four wireless clock synchronization (WCS) methods. (a)--(c) Time-domain TDOA results. (d)--(f) Probability-density distributions. (a, d) Anchors~1\&2. (b, e) Anchors~1\&3. (c, f) Anchors~1\&4.}
  \label{fig:TDOA_meta_comparason}
\end{figure*}

Under the balanced temporal-neighborhood setting ($L_p=3$, $L_d=3$), the four-anchor testbed provides three TDOA anchor pairs for the same typical 50~ms tag slot, enabling an examination of whether AB-Sync improves slot-level ranging across different anchor-pair geometries. The resulting TDOA ranging stability is then compared with the baseline WCS algorithms on the AB-Sync test set. Fig.~\ref{fig:TDOA_meta_comparason} (a)--(c) shows the TDOA values over the test samples using the compared low-overhead WCS methods, while Table~\ref{tab:tdoa_ranging_comparison} reports the quantitative STD.V results for all baselines. In these three figures, a more stable TDOA sequence indicates a superior WCS algorithm, since TDOA stability not only implies consistent range-difference measurements but is also directly related to positioning stability. As shown in Fig.~\ref{fig:TDOA_meta_comparason} (a)--(c), AB-Sync produces more compact TDOA sequences around the ground-truth values than the prediction-based baselines. Table~\ref{tab:tdoa_ranging_comparison} further quantifies this trend: AB-Sync reduces the average STD.V from 7.19~cm with Deferred to 6.48~cm. The gain is observed across the three anchor pairs, indicating that the selected slot-level correction improves TDOA stability across different anchor-pair geometries.

Fig.~\ref{fig:TDOA_meta_comparason} (d)--(f) further show the probability-density distributions of the TDOA values. AB-Sync yields the narrowest distribution among the compared methods, confirming the reduced TDOA dispersion. Deferred+3S-KF remains the strongest non-learning baseline, but AB-Sync further reduces the TDOA STD.V by 12.53\%, 8.51\%, and 1.60\% for Anchors~1\&2, Anchors~1\&3, and Anchors~1\&4, respectively. These gains also follow the anchor-pair geometry. Lyu et al.~\cite{Defer-3s-kf} report that anchor/tag spatial distributions affect TDOA magnitudes and error levels, and that WCS gains become smaller when the residual TDOA error is already low. Consistently, Anchors~1\&2 exhibit the largest fluctuation and gain, whereas Anchors~1\&4 have the smallest fluctuation and show a more limited marginal gain.

\textbf{Cross-pair generalization:} Table~\ref{tab:tdoa_ranging_comparison} further reports the generalization result across anchor pairs. Since Anchors~1\&2 show the largest TDOA fluctuation and AB-Sync gain, AB-Sync (gen., source 1\&2) is trained on this higher-variation pair and directly applied to the other pairs without pair-specific retraining. It reduces STD.V by 12.53\%, 10.31\%, and 3.19\% over Deferred+3S-KF for Anchors~1\&2, Anchors~1\&3, and Anchors~1\&4, respectively. It also outperforms the directly trained AB-Sync models on the two target pairs, reducing STD.V from 5.59 to 5.48~cm and from 5.55 to 5.46~cm. This suggests that training on higher-variation TDOA samples can provide more transferable correction behavior, likely including partial compensation for geometry-dependent measurement fluctuations.

\textbf{Ablation on temporal weighting:} To assess the role of adaptive temporal weighting, we replace the attention module with a compact feed-forward MLP while keeping the same slot-level input, supervision, and training/test split. The MLP still improves over the deferred stage-level correction, reducing STD.V by 4.08\%, 0.88\%, and 0.49\% for Anchors~1\&2, Anchors~1\&3, and Anchors~1\&4, respectively, which confirms that the slot-level temporal context is informative. However, its average STD.V is 7.03~cm, compared with 6.48~cm for AB-Sync, indicating that adaptive temporal weighting extracts stronger correction from the same neighborhood than this non-attention ablation.

\begin{table}[!t]
    \centering
    \caption{TDOA ranging results of AB-Sync, baseline WCS methods, and the MLP ablation (STD.V, cm).}
    \label{tab:tdoa_ranging_comparison}
    \renewcommand{\arraystretch}{1.12}
    \resizebox{\linewidth}{!}{%
    \begin{tabular}{lccc}
    \toprule
    Method & Anchors 1\&2 & Anchors 1\&3 & Anchors 1\&4 \\
    \midrule
    Conventional & 15.92 & 9.68 & 6.89 \\
    Conventional+3S-KF & 11.25 & 7.24 & 6.02 \\
    Deferred & 9.69 & 6.21 & 5.66 \\
    Deferred+3S-KF & 9.50 & 6.11 & 5.64 \\
    MLP (abl.) & 9.29 & 6.16 & 5.64 \\
    AB-Sync & \textbf{8.31} & \underline{5.59} & \underline{5.55} \\
    AB-Sync (gen., source 1\&2) & \textbf{8.31} & \textbf{5.48} & \textbf{5.46} \\
    Direct gain vs. Deferred+3S-KF & 12.53\% & 8.51\% & 1.60\% \\
    Gen. gain vs. Deferred+3S-KF & 12.53\% & 10.31\% & 3.19\% \\
    \bottomrule
    \end{tabular}%
    }
\end{table}

\subsection{TDOA Localization Accuracy}\label{subsection:localization}

After confirming the TDOA ranging improvement in Subsection~\ref{subsection experiment tdoa range}, this subsection examines whether the improved TDOA stability translates into two-dimensional localization accuracy. We use the TDOA measurements from the same single-slot experiment and solve the tag position with the closed-form hyperbolic positioning method.

Since Anchors~1, 2, and 3 form a representative room-level localization geometry, their TDOA measurements are used for static 2D positioning. For AB-Sync, the temporal-neighborhood setting follows the selected configuration $L_p=3$ and $L_d=3$. In addition to direct AB-Sync, AB-Sync (gen.) uses the model trained on Anchors~1\&2 to infer the TDOA of Anchors~1\&3, following the cross-pair generalization result in Table~\ref{tab:tdoa_ranging_comparison}.

\begin{table}[h]
    \centering
    \caption{~Static positioning results of different WCS algorithms.}
    \label{table: static_position}
    \scalebox{1}{
    \renewcommand{\arraystretch}{1}
    \begin{tabular}{cccc}
        \toprule
        \textbf{Method} & \textbf{STD.V(cm)}  & \textbf{RMSE(cm)} & \textbf{ME(cm)} \\
        \midrule
        \textbf{Conventional+3S-KF} & 7.04 & 12.77 & 10.65 \\
        \midrule
        \textbf{Deferred} & 5.82 & 10.70 & 8.98 \\
        \midrule
        \textbf{Deferred+3S-KF} & 5.74 & 10.52 & 8.81 \\
        \midrule
        \textbf{AB-Sync} & \underline{4.61} & \underline{8.65} & \underline{7.32} \\
        \midrule
        \textbf{AB-Sync (gen.)} & \textbf{4.57} & \textbf{8.61} & \textbf{7.30} \\
        \bottomrule
    \end{tabular}
    }
\end{table}

Table~\ref{table: static_position} shows that the TDOA-level improvement can be effectively translated into 2D localization improvement. Compared with Deferred+3S-KF, the strongest non-learning baseline in this experiment, direct AB-Sync reduces STD.V, RMSE, and ME by 19.7\%, 17.7\%, and 16.9\%, respectively. AB-Sync (gen.) further achieves 4.57~cm STD.V, 8.61~cm RMSE, and 7.30~cm ME, corresponding to reductions of 20.3\%, 18.1\%, and 17.2\%. The localization-level gain is larger than the corresponding one-dimensional TDOA STD.V reduction because hyperbolic positioning is nonlinear: the final 2D error is jointly shaped by the TDOA error distribution and the anchor geometry. This slight additional gain of AB-Sync (gen.) is consistent with the transferred TDOA improvement observed for Anchors~1\&3 in Subsection~\ref{subsection experiment tdoa range}.

\begin{figure}[!t]
	\centering

    \includegraphics[scale=0.44]{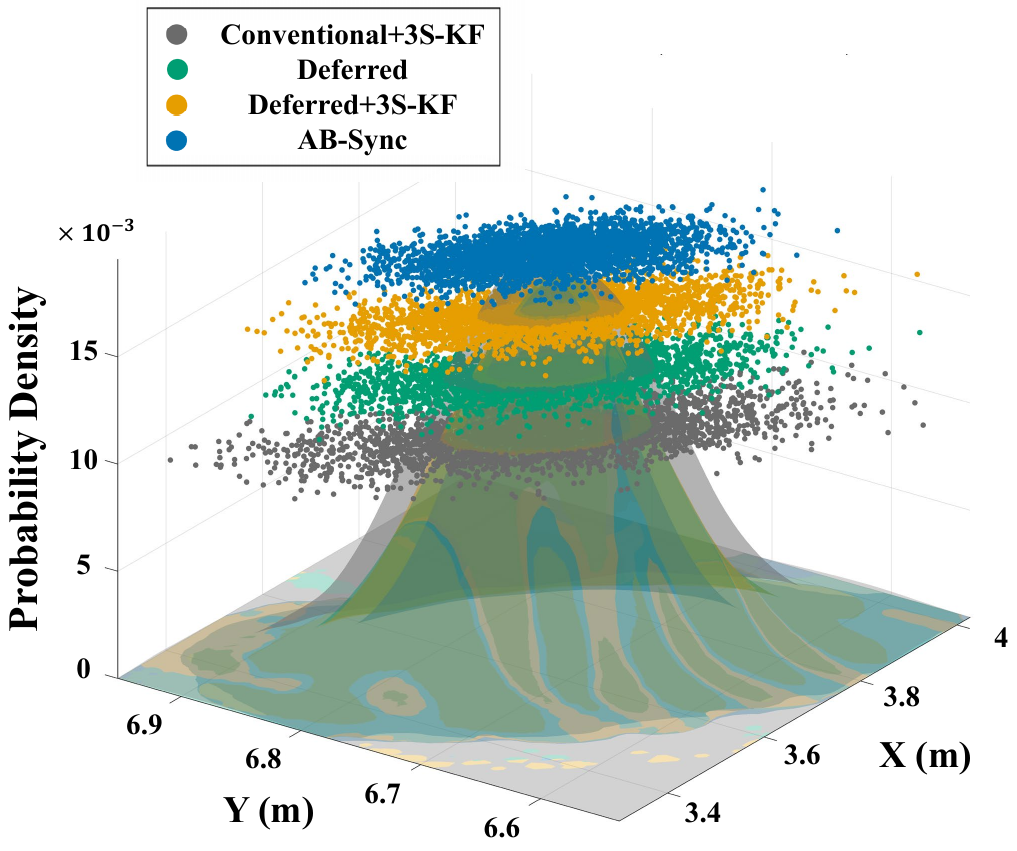}
    
	\caption{Static positioning result distributions of different WCS methods.}
    \label{fig:static position}
\end{figure}

Fig.~\ref{fig:static position} visualizes the resulting localization distributions. AB-Sync produces a more compact point cloud than the baseline WCS methods, which is consistent with the reduced TDOA dispersion in Fig.~\ref{fig:TDOA_meta_comparason}.

\subsection{Multi-Slot Multi-Tag Evaluation}\label{subsection:multi_slot_positioning}

This subsection evaluates AB-Sync in a more practical multi-slot multi-tag setting, where tags occupy representative TDMA slots across one CCP stage and are placed at different field positions. Specifically, three anchors were placed at \((0,0)\), \((4.5,9.3)\), and \((9.3,0)\)~m. Five tags were assigned to different TDMA slots and distributed across the localization area, following the experimental setting described in Table~\ref{tab:ab_sync_training_configuration}. The starts of the five tag slots were placed at 17, 33, 50, 66, and 83~ms after each CCP, covering representative offsets within the 100~ms CCP stage; these slot offsets are also summarized in Table~\ref{tab:multi_slot_static_position_by_slot}. The five tag locations are represented by the empirical localization centers used in the STD.V calculation, namely \((1.8,3.6)\), \((2.1,6.25)\), \((6.4,6.6)\), \((6.75,3.9)\), and \((4.8,4.8)\)~m. For the five tag slots, 22364, 22367, 22345, 22375, and 22350 total available samples were collected, respectively.

Since localization is the end-to-end output of the UWB-TDOA system, the multi-slot experiment is evaluated directly at the localization level, where WCS errors propagate through both TDOA fusion and anchor geometry. Table~\ref{tab:multi_slot_static_position} summarizes the five-slot localization statistics and includes both pair-specific AB-Sync and generalized AB-Sync variants following the cross-pair design in Subsection~\ref{subsection experiment tdoa range}. The reported metrics are averaged over the five tag slots. Compared with Deferred+3S-KF, direct AB-Sync achieves the best average localization result, reducing STD.V from 3.27 to 3.09~cm and lowering RMSE/ME to 5.99/5.11~cm. Table~\ref{tab:multi_slot_static_position_by_slot} further reports the slot-wise STD.V results, showing consistent gains over Deferred+3S-KF across all five TDMA slots. The generalized source-1\&3 setting remains close to direct AB-Sync, achieving 3.10~cm average STD.V versus 3.09~cm. This result indicates that the generalized model preserves most of the localization-level benefit while reducing pair-specific training overhead.

\begin{table}[!t]
\centering
\caption{Multi-slot static positioning results of different WCS algorithms.}
\label{tab:multi_slot_static_position}
\renewcommand{\arraystretch}{1.12}
\resizebox{\columnwidth}{!}{%
\begin{tabular}{lccc}
\toprule
Method & STD.V (cm) & RMSE (cm) & ME (cm) \\
\midrule
Conventional & 4.54 & 9.00 & 7.76 \\
Deferred & 3.35 & 6.51 & 5.57 \\
Conventional+3S-KF & 3.70 & 7.27 & 6.26 \\
Deferred+3S-KF & 3.27 & 6.37 & 5.46 \\
AB-Sync & \textbf{3.09} & \textbf{5.99} & \textbf{5.11} \\
AB-Sync (gen., 1\&2) & 3.12 & 6.06 & 5.18 \\
AB-Sync (gen., 1\&3) & \underline{3.10} & \underline{6.01} & \underline{5.14} \\
\bottomrule
\end{tabular}%
}
\end{table}

\begin{table}[!t]
\centering
\caption{Slot-wise localization STD.V and gain of AB-Sync.}
\label{tab:multi_slot_static_position_by_slot}
\renewcommand{\arraystretch}{1.12}
\resizebox{\columnwidth}{!}{%
\begin{tabular}{lccccc}
\toprule
Metric & Slot~1 & Slot~2 & Slot~3 & Slot~4 & Slot~5 \\
\midrule
Slot offset (ms) & 17 & 33 & 50 & 66 & 83 \\
AB-Sync STD.V (cm) & 3.62 & 2.81 & 3.69 & 1.87 & 3.37 \\
Gain vs. Deferred+3S-KF & 7.1\% & 2.6\% & 3.0\% & 16.2\% & 4.1\% \\
\bottomrule
\end{tabular}%
}
\end{table}

\begin{figure}[!t]
	\centering
    \includegraphics[width=\columnwidth]{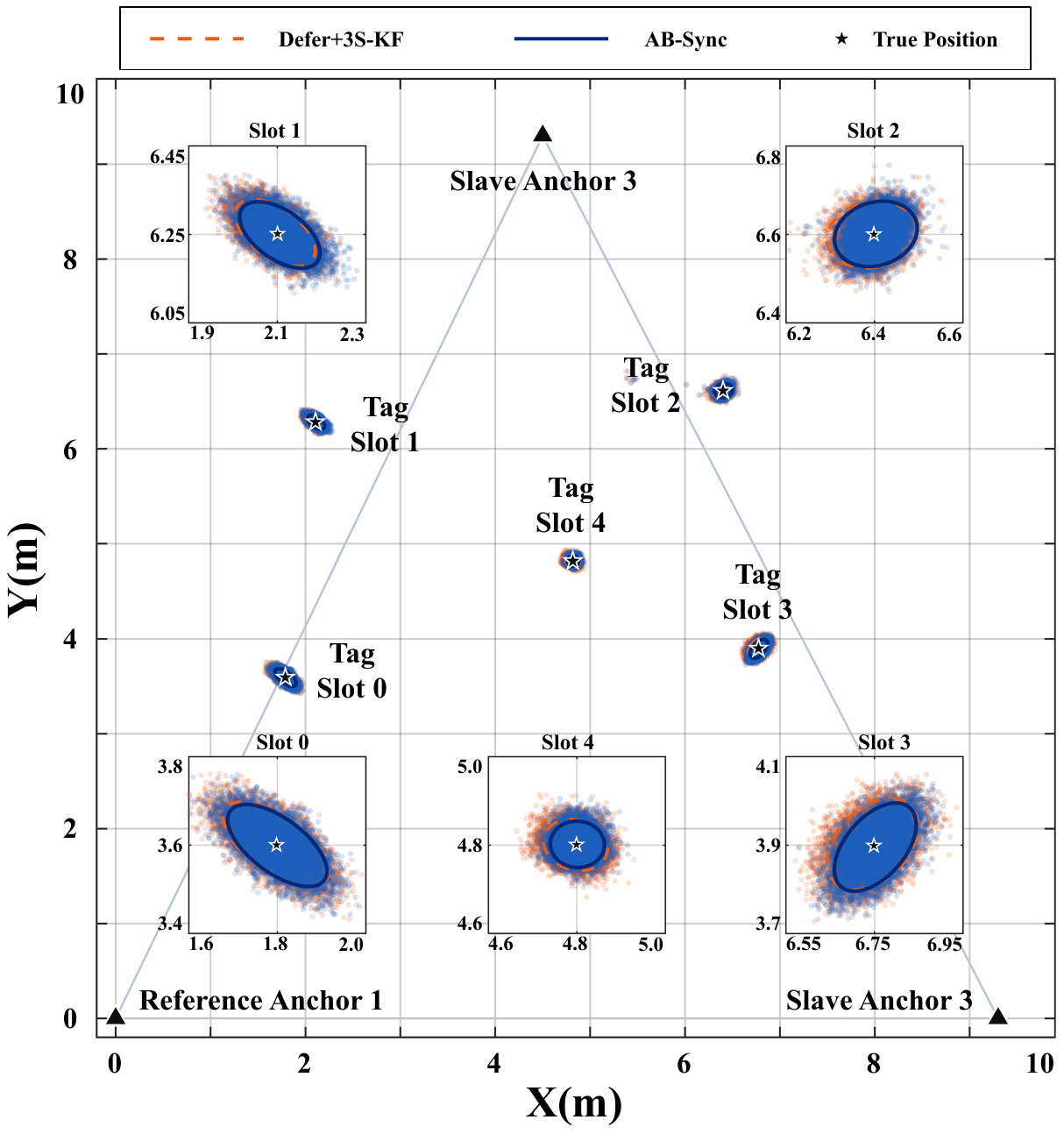}
	\caption{Multi-slot localization result distributions for five static tags transmitted at different TDMA slots. The ellipses denote the two-standard-deviation localization distributions of the prediction-based baseline and AB-Sync.}
    \label{fig:multi_slot_positioning}
\end{figure}

Fig.~\ref{fig:multi_slot_positioning} provides a spatial view of the five-slot localization results. Across the five tag positions, AB-Sync generally produces tighter localization clusters than the prediction-based baseline, which is consistent with the slot-wise STD.V reductions in Table~\ref{tab:multi_slot_static_position_by_slot}. This agreement between the spatial distributions and the quantitative metrics indicates that the slot-level correction remains effective across different TDMA offsets and tag locations.

\section{Related Work}\label{sec:related_work}

UWB-TDOA localization relies on clock synchronization among anchors to map independently measured timestamps onto a common time reference. Anchor clocks can be synchronized through wired links~\cite{pack_10,pack_13} or wireless anchor communication~\cite{ETDOAIRoneway}. Wired synchronization can provide high accuracy, but wireless synchronization is more flexible for practical deployment because it avoids dedicated cabling. Existing wireless clock synchronization (WCS) methods can be broadly categorized as unidirectional~\cite{pack_2,pack_8,pack_14}, bidirectional~\cite{pack_18}, dual-reference~\cite{pack_6}, and distributed synchronization~\cite{pack_9}. Among them, unidirectional WCS is particularly attractive for dense UWB-TDOA networks because only the reference anchor broadcasts Clock Correction Packages (CCPs), leaving more channel time for tag blinks.

A second line of work focuses on estimating and refining the clock state used for timestamp mapping. Linear interpolation and Kalman-filter-based WCS are representative stage-level methods~\cite{Interpolation,2S-KF}. Higher-order filtering further models clock-state evolution, such as 3S-KF and cascaded 3S-KF correction~\cite{3LI-KF_without_a}. Deferred synchronization improves timestamp mapping by waiting until the following CCP becomes available, so the clock state of the stage containing the tag blink can be estimated with lower uncertainty~\cite{2S_KF_deferred,Defer-3s-kf}. Recent learning-based methods also use recurrent models to capture clock-state evolution from synchronization observations~\cite{LSTM}. These methods improve the accuracy of clock-state estimation, but the estimated state is still applied at the CCP-stage or synchronization-interval granularity.

Different from the above stage-level WCS methods, AB-Sync focuses on the stage-to-slot granularity mismatch in TDMA-based UWB-TDOA localization. Rather than designing another stage-level clock-state estimator, AB-Sync targets the residual timing error caused by applying a CCP-stage clock state to a tag blink transmitted in a short TDMA slot. It learns the relationship between neighboring clock-fluctuation observations and the slot-specific clock-speed ratio required for timestamp mapping. In this way, AB-Sync refines UWB-TDOA synchronization from CCP-stage granularity to tag-slot granularity while preserving the low-overhead unidirectional synchronization structure.

\section{Conclusion}\label{sec:conclusion}

This paper presents AB-Sync, an attention-based slot-level clock-deviation correction method for low-overhead UWB-TDOA localization networks. The key idea is to refine the clock state used for TOA mapping from the CCP-stage granularity to the TDMA tag-slot granularity, while retaining the unidirectional one-message synchronization structure. By learning from neighboring clock-fluctuation patterns, AB-Sync improves TDOA ranging stability and localization-level accuracy without introducing additional UWB synchronization messages. The multi-slot experiment further indicates that the learned slot-level correction remains effective across different tag slots and tag locations.

For future work, we will explore interpretable slot-level clock-speed reconstruction from the learned temporal weights, so as to better characterize how attention-based correction differs from model-based Kalman clock evolution within each CCP stage.

\bibliographystyle{IEEEtran}
\bibliography{Reference}
\end{document}